\documentclass[12pt,onecolumn]{article}
\usepackage{graphicx,amssymb,amsmath}

\newcommand\pubnumber{}
\newcommand\pubdate{\today}

\def\myaddress{Department of Physics and the Cyclotron Institute\\
Texas A\&M University, Texas, USA 77843-3636}

\def\Title#1{\begin{center} {\Large #1 } \end{center}}
\def\Author#1{\begin{center}{ \sc #1} \end{center}}
\def\Address#1{\begin{center}{ \it #1} \end{center}}

\newcommand\pubblock{\rightline{\begin{tabular}{l} \pubnumber\\
         \pubdate  \end{tabular}}}
\newenvironment{Abstract}{\begin{quotation}  }{\end{quotation}}
\newenvironment{Presented}{\begin{quotation} \begin{center} 
             PRESENTED AT\end{center}\bigskip 
      \begin{center}\begin{large}}{\end{large}\end{center} \end{quotation}}
\def\Acknowledgements{\bigskip  \bigskip \begin{center} \begin{large}
             \bf ACKNOWLEDGEMENTS \end{large}\end{center}}

\def\beq{\begin{equation}}
\def\eeq#1{\label{#1}\end{equation}}
\def\eeqn{\end{equation}}

\def\beqa{\begin{eqnarray}}
\def\eeqa#1{\label{#1}\end{eqnarray}}
\def\eeqan{\end{eqnarray}}

\let\bar=\overbar

\def\Dslash{\not{\hbox{\kern-4pt $D$}}}
\def\dslash{\not{\hbox{\kern-2pt $\del$}}}

\def\msb{{\bar{\ssstyle M \kern -1pt S}}}

\newcommand{\tsups}[1]{\textsuperscript{#1}}

\begin{document}
\begin{titlepage}
\pubblock

\vfill
\Title{Nuclear $\beta$-decay measurements and $|V_{ud}|$}
\vfill
\Author{Dan Melconian}
\Address{\myaddress}
\vfill
\begin{Abstract}
  Some recent work in nuclear $\beta$ decay related to the value of 
  $|V_{ud}|$ is described along with some near-term goals for future 
  measurements.
\end{Abstract}
\vfill
\begin{Presented}
  The 6th International Workshop on the CKM Unitarity Triangle\\
  University of Warwick, UK, September 6--10, 2010
\end{Presented}
\vfill
\end{titlepage}
\def\thefootnote{\fnsymbol{footnote}}
\setcounter{footnote}{0}

\section[Isospin T=1 Superallowed Decays]{Isospin \boldmath$T\!=\!1$ Superallowed Decays}
The comparative half-lives -- or $ft$ values -- of the $\beta$ decay of 
nuclei, specifically $I^\pi=0^+\rightarrow0^+$ isospin \mbox{$T=1$} pure Fermi 
transitions, provide the most precise value of $|V_{ud}|$ to 
date~\cite{hardy-PRC79-2009}.  Nevertheless, a number of groups 
at a number of facilities continue to improve/confirm the experimental inputs 
going into the determination of $|V_{ud}|$:  the decay energy for the 
phase-space factor, $f$, as well as the half-life and $\beta$ branch for $t$.  
Most notably, recent work on mass measurements using Penning traps have 
improved the $Q_\mathrm{EC}$ values of some of these transitions.  In 
2005 using the Canadian Penning trap, the group at Argonne 
National Laboratory measured the value for \tsups{46}V and found a large 
discrepancy~\cite{savard-PRL95-2005} from earlier measurements based on 
$(^3\mathrm{He},\mathrm{t})$ reactions~\cite{vonach-NuclPhysA278-1977}.  
This result was confirmed using the JYFLTRAP Penning trap at the University of 
Jyv\"askyl\"a~\cite{eronen-PRL97-2006}, as well as by a new 
$(^3\mathrm{He},\mathrm{t})$ reaction 
measurement~\cite{faestermann-EPJA42-2009} using a very similar set-up as was 
originally done.  Upon further investigation, 
Penning trap measurements also found discrepancies in the cases of 
\tsups{42}Sc, \tsups{50}Mn and \tsups{54}Co~\cite{eronen-PRL100-2008} and 
led to Penning trap campaigns at ANL, Jyv\"askyl\"a, ISOLTRAP at 
ISOLDE~\cite{george-EPL82-2008} and LEBIT at the National Superconducting 
Labotory~\cite{ringle-PRL96-2006} to check and/or improve the 
$Q_\mathrm{EC}$ values of all the \mbox{$T=1$} superallowed transitions used to 
determine $|V_{ud}|$.  Other recent Penning trap measurements which have 
improved the $Q_\mathrm{EC}$ value (but don't show a deviation from earlier 
values) include 
\tsups{18}Ne, \tsups{22}Mg, \tsups{26}Al$^m$, \tsups{26}Si, \tsups{30}S, 
\tsups{34}Ar, \tsups{34}Cl, \tsups{38}Ca, \tsups{38}K$^m$, 
\tsups{42}Ti, \tsups{62}Ga, \tsups{66}As and \tsups{74}Rb.

The value of $|V_{ud}|$ is calculated based on the average of the 
``$\mathcal{F}t$ values,'' which are the observed $ft$ values corrected for 
small radiative and nuclear-structure dependent effects.  The latest review 
of the superallowed determination of $|V_{ud}|$ by Hardy and 
Towner~\cite{hardy-PRC79-2009}, which describes this process in greater 
detail, includes many of these mass measurements described above and 
discusses the implications.  They find
\begin{align}
  |V_\mathrm{ud}| = 0.97425(21)_\mathrm{theor}(8)_\mathrm{exp} 
\end{align}
The transition-independent part of the radiative correction (common to 
\emph{any} evaluation of $|V_{ud}|$) is the dominant uncertainty, contributing 
$18\times10^{-5}$ to the error budget.  Note that this is twice as large as 
the uncertainty due to experiment.

While the \mbox{$T=1$} superallowed decays are the most precise, they are not 
the only manner in which $|V_{ud}|$ can be determined.  Furthermore, as 
discussed in Ref.~\cite{hardy-PRC79-2009} (see also I.S.\ Towner's article in 
these proceedings), there appears to be some model-dependence to the 
theoretical corrections used to calculate the $\mathcal{F}t$ values from the 
observed $ft$ values.  It is therefore important to 
compliment the value of $|V_{ud}|$ obtained from \mbox{$T=1$} transitions with 
other methods; even if these methods are not as precise, they can test the 
theoretical corrections used and provide new values of $|V_{ud}|$ which have 
completely different systematic uncertainties.  Agreement between the 
complementary methods would help reduce concerns about the theoretical 
corrections that are applied.

\begin{figure}[htb]
  \centering
  \includegraphics[width=0.75\textwidth]{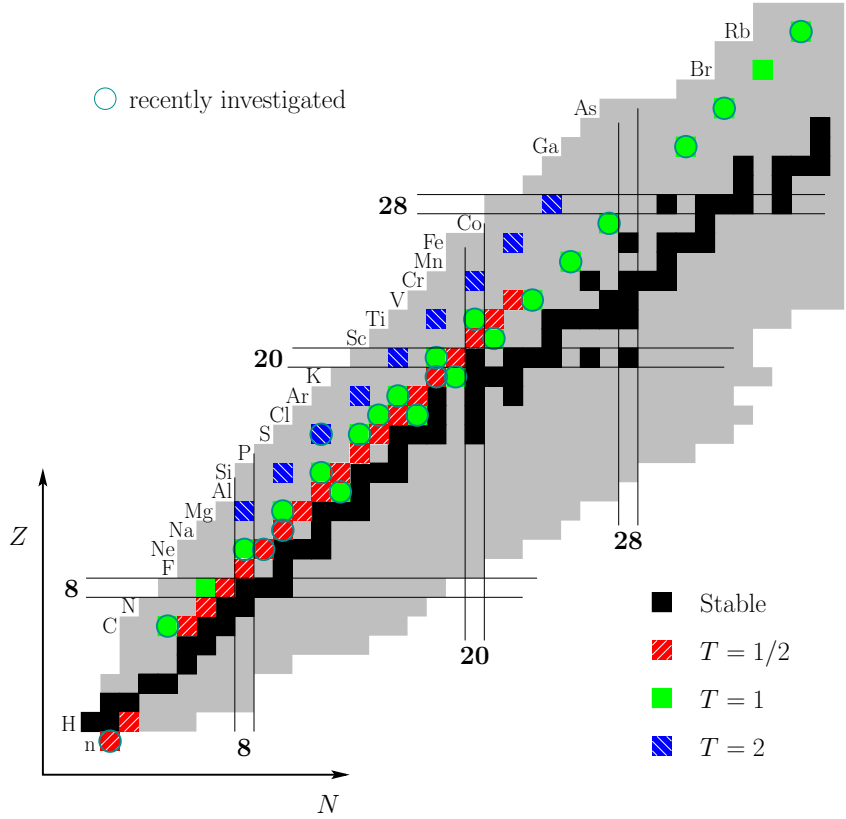}
  \caption{(Colour online) Chart of the nuclides showing superallowed 
    ($T=1,2$) and mirror transitions ($T=1/2$) relevant to determining the 
    value of $|V_{ud}|$.}
  \label{fig:melconian:Q-values}
\end{figure}

Figure~\ref{fig:melconian:Q-values} shows the chart of the nuclides, 
highlighting the radioactive nuclei that are of interest to determining the 
value of $|V_{ud}|$.  Within the \mbox{$T=1$} transitions, a group at TRIUMF 
recently measured the branching ratio of \tsups{62}Ga and interpreted their 
results as a test of the isospin-mixing corrections~\cite{hyland-PRL97-2006}.  
They found that calculations overestimate the effect, thus indicating a bigger 
shell-model space is required in this and other heavy ($A\geqslant62$) nuclei. 
Other work at TRIUMF that is currently in progress includes improving the 
half-life of \tsups{26}Al$^m$~\cite{finlay-privcomm} and using a novel 
charge-breeding electron ion beam trap with a Penning trap to measure the 
mass of \tsups{74}Rb~\cite{dilling-privcomm}.  At Jyv\"askyl\"a, their 
program continues with half-life measurements of \tsups{29}P, \tsups{31}S 
(also it's branching ratio) and  \tsups{39}Ca.  The Cyclotron Institute at 
Texas A\&M University is analyzing data on half-life measurements of 
\tsups{10}C and \tsups{26}Si, and a recently completed analysis of the 
branching ratio in \tsups{32}Cl validates the shell-model prediction 
of isospin-mixing in the $s,d$-shell~\cite{32Cl}.  All of these programs 
combined will reduce uncertainties in the $\mathcal{F}t$ values of 
\mbox{$T=1$} superallowed transitions, allowing a more reliable value of 
$|V_{ud}|$ to be determined. 

\section[T=2 Superallowed Decays]{\boldmath$T\!=\!2$ Superallowed Decays}
Hardy and Towner~\cite{hardy-PRC79-2009} have compared their Woods-Saxon 
potential based shell-model calculations of isospin-mixing corrections to 
one based on Hartree-Fock eigenfunctions, and they find indications of a 
dependence on isospin:  the average difference in the predicted correction 
between these two models for \mbox{$T_3=-1$} cases is larger than for the 
\mbox{$T_3=0$} transitions.  By going to higher $T$ multiplets, a comparison 
of these models may be tested with enhanced sensitivity.  Recently, a 
measurement of the isospin-mixing in \tsups{32}Ar was 
made~\cite{bhattacharya-PRC77-2008} by improving the branching ratio of this 
isospin \mbox{$T=2$} superallowed decay.  This proton-rich nucleus $\beta^+$ 
decays to a proton-unbound state in \tsups{32}Cl, thus requiring a measure of 
the proton and $\gamma$ branches from the $0^+$ excited state in \tsups{32}Cl. 
The total uncertainty in the $\mathcal{F}t$ value of this decay remains 
dominated by the branching ratio, and we have plans to improve it using a 
large-bore, open-geometry cyclindrical Penning trap to be built at the 
Cyclotron Institute, Texas A\&M University.  Once this technique is proven 
for \tsups{32}Ar, there are a number of other \mbox{$T=2$} nuclei which can 
be produced at the Cyclotron Institute and decay in a similar manner:  
\tsups{20}Mg, \tsups{24}Si, \tsups{28}S, \tsups{36}Ca and \tsups{40}Ti.  Using 
a \tsups{3}He target and standard projectile beams, LISE calculations indicate 
rates into the Penning trap ranging from $1\times10^3$/s (\tsups{20}Mg) up to 
a few times $10^6$/s (\tsups{32}Ar, \tsups{36}Ca and \tsups{40}Ti).  Even the 
lowest rate \tsups{20}Mg is enough to mount a precision experiment due to the 
extremely low backgrounds of the Penning trap.  Once the $ft$ values of these 
nuclei are measured precisely, there will be six new superallowed decays which 
can help test models of isospin-mixing.  Given acceptance of a model for these 
proton-rich cases, one could add them to the list of transitions used to 
extract $|V_{ud}|$; this would have a higher impact on reducing the uncertainty 
in $|V_{ud}|$ than improving already very precise measurements in the 
\mbox{$T=1$} cases.

\section[T=1/2 Transitions]{\boldmath$T\!=\!1/2$ Transitions}
\subsection{Neutron decay}
The neutron is theoretically the simplest nuclear system from which one may 
deduce $|V_{ud}|$.  Unfortunately, it is difficult to measure accurately its 
long lifetime and, since it has a Gamow-Teller component to its decay, 
one must also measure an angular correlation parameter to extract $|V_{ud}|$.  
The contribution of B.~M\"arkisch in these proceedings discusses the neutron 
in depth, so we will not discuss it further here.

\subsection{Mirror transitions}
As pointed out recently by Navialiat-Cuncic and 
Severijns~\cite{naviliat-PRL-2009}, the $\mathcal{F}t$ values of \mbox{$T=1/2$} 
$\beta$ decays between isobaric analogue states may be used as a new avenue 
to deduce $|V_{ud}|$.  With the same theoretical treatment used to calculate 
corrections to the \mbox{$T=1$} superallowed decays applied to these mirror 
transitions in Ref.~\cite{severijns-PRC78-2008}, one may survey the data 
for \mbox{$T=1/2$} decays and extract an independent value for $|V_{ud}|$.  
As with the neutron, a correlation parameter is required in addition to the 
$\mathcal{F}t$ value in order to determine $|V_{ud}|$.  Specifically, 
the master equation used to determine $|V_{ud}|$ from mirror transitions is:
\begin{align}
  |V_{ud}|^2 = \frac{5831.3\pm2.3~\mathrm{s}}{\mathcal{F}t^\mathrm{mirror}
    \left(1+\frac{f_A}{f_V}\rho^2\right)},\label{eq:melconian-master-equation}
\end{align}
where $f_A$/$f_V$ is the ratio of statistical rate functions for axial/vector
currents (which ranges from 0.988--1.04, but typically only differ from unity 
by less than 2\%), and $\rho=C_AM_{GT}/C_VM_F$ is the ratio of Gamow-Teller 
to Fermi strengths for the decay.  The correlation typically used to 
determine $\rho$ is the $\beta$ asymmetry (much like that correlation is 
used in neutron decay to determine $\lambda$), however other correlations 
can and have been used (e.g.~the neutrino asymmetry and the $\beta-\nu$ 
correlation).

\begin{figure}[htb]
  \centering
  \includegraphics[angle=90,width=0.75\textwidth]{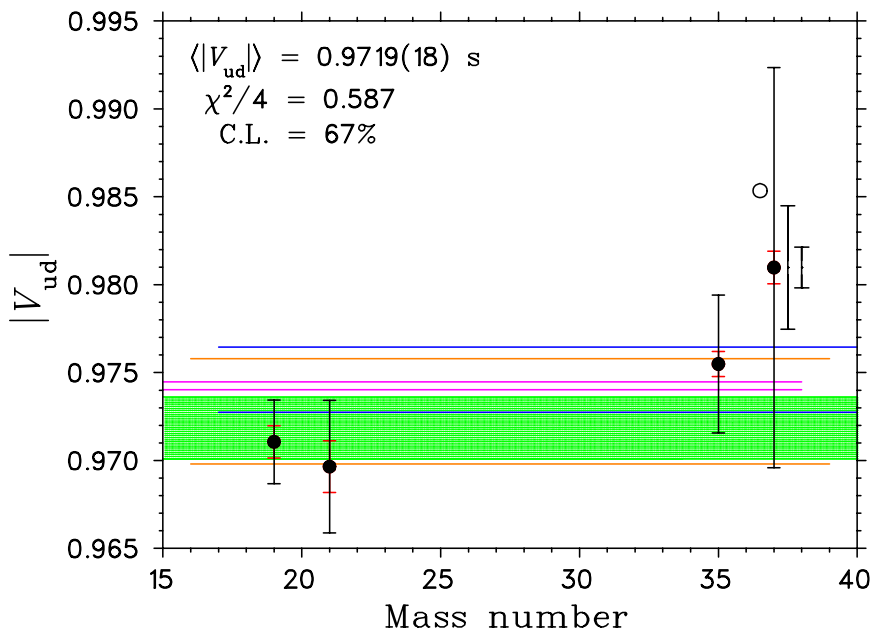}
  \caption{(Colour online) Measurements of $|V_{ud}|$ from mirror transitions 
    (adapted from Ref.~\cite{naviliat-PRL-2009}).  The average value from 
    these five cases where a correlation has been measured yields 
    $|V_{ud}|^\mathrm{mirror}=0.9719(18)$, which is already of the same 
    precision as the neutron (using the PDG values).  The inner red error 
    bars show the statistical uncertainty in each measurement of $|V_{ud}|$ 
    \emph{excluding} the uncertainty in $\rho$.  The dashed green line 
    shows the $1\sigma$ allowed value of $|V_{ud}|$.  }
  \label{fig:melconian:mirror-V_ud}
\end{figure}

To date, only five cases of \mbox{$T=1/2$} mirror transitions have their 
$\mathcal{F}t$ value and a correlation parameter measured to allow a 
determination of $|V_{ud}|$:  \tsups{19}Ne, \tsups{21}Na, \tsups{29}P, 
\tsups{35}Ar and \tsups{37}K.  A plot of these results is shown in 
Fig.~\ref{fig:melconian:mirror-V_ud}.  To help indicate that the correlation 
parameter measurements are currently limiting the extraction of $|V_{ud}|$ 
(rather than the $\mathcal{F}t$ values or theoretical uncertainties), the 
inner error bars show what the uncertainty would be if $\rho$ was known to 
perfect precision; clearly it is this avenue experimentalists must pursue if 
we are to improve the value of $|V_{ud}|$ from mirror transitions.  

Although not relevant to $|V_{ud}|$, it should be noted that improved 
measurements of the $\mathcal{F}t$ values of these and other mirror 
transitions is interesting for other Standard Model tests.  By measuring 
the comparitive half-life to a greater precision, one may \emph{assume} 
that the \mbox{$T=1$} average $\mathcal{F}t$ value is correct and use that to 
\emph{deduce} $\rho$ for the mirror transitions (\emph{i.e.}~re-arrange 
Eq.~\eqref{eq:melconian-master-equation} and solve for $\rho$).  One can then 
calculate the Standard Model prediction of the correlation parameters and 
compare them to experimentally observed values; new physics such as 
right-handed currents, second-class currents, leptoquarks, etc., would 
affect the value of the correlation parameters that one may be able to 
detect.  The interested reader is referred to Refs.~\cite{severijns-PRC78-2008} 
and~\cite{profumo-PRD75-2007}.

As an example of recent work with these nuclei, our group at the Cyclotron 
Institute (and also at Kernfysisch Versneller Instituut in Groningen) is 
completing analysis of a \tsups{37}K lifetime measurement which will reduce 
the uncertainty in $\rho$ by about an order of magnitude.  This will improve 
the value of $\rho$ if one assumes that the average $\mathcal{F}t$ values of 
\mbox{$T=1$} decays is correct, allowing more definite predictions of the 
correlation parameters; however, it does essentially nothing to improve the 
measured $|V_{ud}|$ from \tsups{37}K because the 3\% measurement of the neutrino 
asymmetry parameter~\cite{melconian-PLB-2007} totally dominates the measured 
value of its $\mathcal{F}t$ value.  We are planning to measure the $\beta$ 
asymmetry parameter from this decay to $\lesssim0.5\%$ using the 
magneto-optical trap and high production rates of \tsups{37}K at TRIUMF by 
early 2012.  This measurement combined with the improved lifetime \emph{will} 
greatly reduce the total uncertainty in the \tsups{37}K point of 
Fig.~\ref{fig:melconian:mirror-V_ud} and therefore improve the value of 
$|V_{ud}|$ from mirror decays as a complement to the value obtained from the 
\mbox{$T=1$} superallowed transitions.

\section{Summary and Outlook}
The experimental and theoretical investigations into the value of $|V_{ud}|$ 
from the nuclear physics community continues to be a vibrant field.  Ever 
more precision measurements of the traditional \mbox{$T=1$} superallowed decays 
are being made; re-measurments of the masses using Penning trap mass 
spectrometers have led to discovering a small bias in older measurements, 
but recently most emphasis has been on trying to \emph{measure} the 
isospin-mixing corrections in nuclei to test the theoretical corrections 
used to extract $|V_{ud}|$.

There are very recent measurements of the neutron asymmetry from the PERKEO 
and UCNA collaborations, with more experiments planned to finally determine 
the value of $\lambda$.  In addition, a number of lifetime measurements are 
planned to resolve the outstanding $8\sigma$ discrepancy in previous 
neutron lifetimes with the recent result of Serebrov \emph{et al.}  Once 
all these experiments produce results, we can expect the value of $|V_{ud}|$ 
from neutron decay to improve dramatically and meaningfully compliment that 
of the \mbox{$T=1$} superallowed decays.

Finally, a new avenue of other \mbox{$T=1/2$} mirror transitions has recently 
become available as another nuclear measurement of $|V_{ud}|$ now that the 
isospin and radiative corrections for these nuclei have been calculated.  
Already with even just a few cases measured to $\lesssim0.5\%$, the precision 
of $|V_{ud}|$ from these decays is at the same level of precision as the 
neutron.  The nuclear physics community will also continue to measure 
correlation parameters in these decays which will provide yet another 
important, complimentary measurement of $|V_{ud}|$.


\Acknowledgements
I am grateful to T.\ Eronen, O.\ Naviliat-Cuncic, J.\ Dilling and P.\ Finlay 
for useful discussions.  I would also like to acknowledge the work and efforts 
of my colleagues at Texas A\&M, Los Alamos National Laboratory (UCNA) and 
TRIUMF (TRINAT).

\end{document}